\documentclass[twocolumn,prl,showpacs,floatfix]{revtex4}
\usepackage{amsmath}
\usepackage{amssymb}
\usepackage{bm}
\usepackage{graphicx}
\usepackage{verbatim}
\usepackage{color}
\usepackage{subfigure}

\begin{document}

\title{Dynamical Self-Quenching of Spin Pumping into Double Quantum Dots}

\begin{abstract}
Nuclear spin polarization can be pumped into spin-blocked quantum dots by multiple Landau-Zener passages through singlet-triplet anticrossings. By numerical simulations of realistic systems with $10^7$ nuclear spins during $10^5$ sweeps, we uncover a mechanism of dynamical self-quenching which results in a fast saturation of the nuclear polarization under stationary pumping. This is caused by screening the random field of the nuclear spins. For moderate spin-orbit coupling, self-quenching persists but its  patterns are modified. Our finding explains low polarization levels achieved  experimentally and calls for developing new protocols that break the self-quenching limitations.
\end{abstract}

\author{Arne Brataas$^{1}$ and Emmanuel I. Rashba$^{2}$}
\affiliation{Department of Physics, Norwegian University of Science and Technology,
NO-7491 Trondheim, Norway\\
Department of Physics, Harvard University, Cambridge, Massachusetts 02138,
USA}

\maketitle

Double quantum dots (DQDs) are promising platforms for spintronics \cite{Zutic} and quantum computing \cite{LDV,Levy:prl02,Awsch02,Hanson:rmp07}. For qubits encoded in singlet ($S$) and triplet ($T$) states of spin blockaded DQDs \cite{Ono2002}, the hyperfine coupling of electron spins to the nuclear spin reservoir is critical. Although electron spin relaxation caused by this coupling is destructive, a properly controlled nuclear polarization is an efficient tool  for performing rotations of $S$-$T_0$ qubits \cite{Petta2005,Shulman2012}; $T_0$ is the zero component of the electron spin triplet ($T_0,T_\pm$). A widely discussed approach for pumping nuclear spin polarization into a DQD is based on multiple Landau-Zener (LZ) passages across the $S-T_+$ anticrossing \cite{Ramon:prb07,Gullans:prl10,Rudner:prb10b,Burkard,BR2011} ($T_+$ is the lowest energy component of the triplet $T$ in GaAs).  In the absence of spin-orbit (SO) coupling, angular momentum conservation requires that each transformation of the $S$ state into the $T_+$ state is accompanied by the net transfer of one quantum unit of angular momentum to the nuclear subsystem. Multiple $S\rightarrow T_+$ passages increase the nuclear spin polarization, but experimental data show that spin pumping typically saturates at a surprisingly low level of about 1\% \cite{Reilly:2010}, and the origin of this puzzling behavior remains unknown. Higher levels of the nuclear polarization differences (``gradients") between the two dots were only achieved by using feedback loop schemes \cite{Bluhm:prl10}.  

In this Letter, we show that nuclear spin pumping produces dynamical screening of {\it both} the SO coupling and the random hyperfine (Overhauser) field controlling the width of the $S-T_+$ anticrossing as well as the efficiency of spin pumping, while the detailed patterns differ.  The screening results in one of the nuclear spin configurations with a vanishing anticrossing width, $v\rightarrow0$. Hence, the probability of the $S\rightarrow T_+$ transition and the angular momentum transfer inevitably vanish, resulting in quenching of spin pumping. We call this  \emph{dynamical self-quenching} of spin pumping into double quantum dots borrowing the word ``self-" from the theory of polarons, where self-trapping implies a joint evolution of the electron and phonon subsystems \cite{polaron}.

As applied to hyperfine coupled systems, this conclusion appears to agree with the concept of dark states envisioned in Ref.~\cite{TaylorIL:PRL2003} and further discussed in Ref.~\cite{Gullans:prl10}; the latter paper is mostly concerned with the building of gradient fields. However, the existence of dark states has no direct experimental confirmation yet. On the theoretical side, the patterns of highly-nonlinear coupled electron-nuclear dynamics that might bring systems including millions of nuclear spins into such states remain unclear. To resolve the problem, we performed large scale numerical simulations for realistic systems. Our procedure (i) evaluates the coherent precession of the coupled electron spin and about $10^7$ nuclear spins subject to an external magnetic field during a large number (up to $10^5$) of LZ sweeps through the $S-T_+$ anticrossing and (ii) computes the building of the nuclear polarization during each LZ sweep.
The calculations unveiled the gross features of the self-quenching process. Among our results, the following are of special importance: 
(i) self-quenching sets in under generic conditions, (ii) spin-orbit interaction is dynamically screened despite the violation of the angular momentum conservation,  (iii)
durations of $S-T_+$ pulses have a critical effect on the quenching dynamics, and (iv)
dynamical screening is robust with respect to moderate noise levels.

We consider two electrons in a double quantum dot that can be in singlet  or triplet  states and represent the orbital part of the wave function as $\psi_S(\bf{r}_1,\bf{r}_2)$
or $\psi_T(\bf{r}_1,\bf{r}_2)$.  
The electrons are coupled to the nuclear spins via the hyperfine coupling Hamiltonian 
\begin{equation}
H_{hf} = V_s \sum_{\lambda} A_\lambda \sum_{j \in \lambda} \sum_{m=1,2} {\bf I}_{j \lambda} \cdot {\bf s}(m)\delta({\bf R}_{j\lambda} - {\bf r}_\text{m}) \, , 
\label{Hhf}
\end{equation}
where ${\bf s}(m)= \mbox{\boldmath{$\sigma$}}(m)/2$ are the electron spin operators in terms of the Pauli matrices $\mbox{\boldmath{$\sigma$}}$, $m=1,2$ enumerates electrons, ${\bf r}_m$
are electron coordinates, ${\bf I}_{j\lambda}$ are nuclear spins, $\lambda$ enumerates nuclear species and $j$ lattice sites ${\bf R}_{j\lambda}$, $A_\lambda$ are hyperfine coupling constants for the species $\lambda$, and $V_s$ is a volume per unit cell.  We consider GaAs that has three spin $I=3/2$ nuclear species $^{69}$Ga, $^{71}$Ga, and $^{75}$As. All GaAs parameter values are  known  \cite{Schliemann:jpcm03,Hanson:rmp07,Taylor:prb07} and listed in Ref.~\cite{Supplemental}. The electrons and nuclear spins are subject to an external magnetic field which is aligned along the $z$-direction.

Before presenting our numerical results, we review how electronic Landau-Zener sweeps influence nuclear spins \cite{BR2011}. The hyperfine interaction of Eq.~(\ref{Hhf}) couples two electrons to a large number of nuclear spins. We account for the effect of this interaction by using a semi-classical Born-Oppenheimer aproach so that the slow nuclear spins produce a coupling between the singlet and triplet electronic states. In turn, the  nuclear spins are driven by the electron dynamics controlled by time-dependent gate voltage variations.  The variations causing Landau-Zener transitions occur in the interval $-T_\text{LZ} \leq t \leq  T_\text{LZ}$  and they are repeated many times after waiting for a time $T_\text{w}$, where $T_\text{w} \gg T_\text{LZ}$. During the waiting time $T_\text{w}$, the nuclear spins are only affected by the external magnetic field and not by the electrons. During the LZ sweeps, the voltage changes also induce relative shifts of the electron singlet and triplet energy levels driving passages of the system through the $S-T_+$ anti-crossing,  Fig.\ \ref{GaAsmultispecie12345}(a).  Restricting  the discussion to its vicinity and disregarding contributions of the ($T_0,T_-$) spectrum branches, the coupled equations for singlet and triplet amplitudes $c_S(t)$ and $c_{T_+}(t)$ are ($\hbar=1$) 
\begin{equation}
i\partial_t\left(\begin{array}{c}c_S\\c_{T_+}\end{array}\right)=\left(\begin{array}{cc}\epsilon_S&v^+\\v^-&\epsilon_{T_+}-\eta\end{array}\right)\left(\begin{array}{c}c_S\\c_{T_+}\end{array}\right),
\label{Hst}
\end{equation}
where $\epsilon_S(t)$ and $\epsilon_{T_+}(t)$ are electronic energies controlled by the gates. The off-diagonal matrix elements of Eq.\ (\ref{Hst}), $v^\pm=v_\text{n}^\pm+v_\text{so}^\pm$, include contributions from the nuclear spins generated by the hyperfine coupling 
\begin{equation}
v_\text{n}^{\pm}=V_{s}\sum_{\lambda}A_{\lambda}\sum_{j\in\lambda}\rho_{j\lambda}(I_{j\lambda}^{x}\pm iI_{j\lambda}^{y})/\sqrt{2}
\label{eq5}
\end{equation}
and from the spin-orbit coupling $v_\text{so}^{\pm}$ \cite{Rudner:prb10a}. The diagonal contribution $\eta=\eta_Z+\eta_\text{n}$  includes the Zeeman energy of the $T_+$ state in the external magnetic field ${\bf B}=B{\hat{\bf z}}$ and the 
Overhauser field of the nuclear polarization, 
 $\eta_\text{n}=-V_{s}\sum_{\lambda}A_{\lambda}\sum_{j\in\lambda}\zeta_{j\lambda}I_{j\lambda}^{z}$ which
is defined solely by the coupling of the triplet component of the electron wave function to the longitudinal nuclear spin polarization. 
The singlet-triplet coupling constants are $\rho_{j\lambda}=\int d\mathbf{r}\psi_{S}^{*}(\mathbf{r},\mathbf{R}_{j\lambda})\psi_{T}(\mathbf{r},\mathbf{R}_{j\lambda})$
and the
coupling constants in the $T_{+}$ state are $\zeta_{j\lambda}=\int d\mathbf{r}\left|\psi_{T}(\mathbf{r},\mathbf{R}_{j\lambda})\right|^2$.

Between the Landau-Zener cycles, during every waiting period of duration $T_\text{w}$, the nuclear spins precess in the external magnetic field.
It is assumed that the time-scale $T_\text{LZ}$ for the LZ transition (induced by  rapid changes in the gate voltages) is much shorter than the nuclear precession times $t_{pr}$ for the species in the external magnetic field, $t_\text{71As}$, $t_\text{69Ga}$, and $t_\text{71Ga}$, respectively.  Because of the large number of nuclei in the dot, $N\sim 10^6-10^7$, the changes $\triangle\mathbf{I}_{j\lambda}$ acquired by 
individual spins during a single sweep are minor. We evaluate them by integrating the equations of coherent nuclear dynamics $d{\bf I}_j/dt=\mbox{\boldmath$\Delta$}_j\times{\bf I}_j$ over the Landau-Zener transition time 
$-T_\text{LZ} \leq t \leq T_\text{LZ}$. Here $\mbox{\boldmath$\Delta$}_j$ are the Knight fields following from Eq.~(\ref{Hhf}). Finally, $\triangle\mathbf{I}_{j\lambda}=\boldsymbol{\Gamma}_{j\lambda}\times\mathbf{I}_{j\lambda}$, where
\begin{subequations}
\begin{align}
\Gamma_{j\lambda}^{(x)} & =-V_{s}A_{\lambda}\rho_{j\lambda}\left(Pv_{y}+Qv_{x}\right)/(2v^{2}) , \\
\Gamma_{j\lambda}^{(y)} & =V_{s}A_{\lambda}\rho_{j\lambda}\left(Pv_{x}-Qv_{y}\right)/(2v^{2}) , \\
\Gamma_{j\lambda}^{z} & =V_{s}A_{\lambda}\zeta_{j\lambda}R/(2v) ,
\end{align}
\label{spinchanges}
\end{subequations}
with $v=\left|v^\pm\right|=\sqrt{{(v_{x}^{2}+v_{y}^{2})}/2}$ and $v^\pm=(v_x\pm i v_y)/\sqrt{2}$. In these expressions, $0\leq P\leq1$ is the $S$-$T_{+}$ transition
probability 
and an unbounded real number $Q$ is the shake-up parameter defined as \cite{BR2011} 
\begin{equation}
P+iQ=-i2v^{-}\int_{-T_\text{LZ}}^{T_\text {LZ}} dt~ c_{S}(t)c_{T_{+}}^{*}(t)
\label{eq:PandQ}
\end{equation}
in terms of the singlet (triplet) amplitudes,
and $R=2v\intop_{-T_\text {LZ}}^{T_\text {LZ}}dt\left|c_{T_{+}}(t)\right|^{2}$
accounts for the Knight shift 
due to the electron spin in the triplet $T_{+}$ state during the 
Landau-Zener transition.

It follows from Eq.~\ref{Hst} and Eq.~\ref{eq:PandQ} that $P$ is equal to the change of the occupation of the singlet state $\vert c_S(t)\vert^2$ during the sweep and therefore coincides with the LZ transition probability. 
The change $\Delta I_z$ in the total longitudinal magnetization $I_z$ equals
\begin{equation}
\Delta I_z=-\frac{P}{2v^2}(v^-v_\text{n}^++v^+v_\text{n}^-)-i\frac{Q}{2v^2}(v^-v_\text{n}^+-v^+v_\text{n}^-).
\label{DeltaIz}
\end{equation}
When $v_\text{so}^\pm=0$, the second term in Eq.~\ref{DeltaIz} vanishes and $\Delta I_z=-P$ as required by the conservation of the angular momentum. In the same scenario,
$Q\neq0$ and describes the angular momentum transfer inside the DQD due to the shakeup processes induced by the LZ pulses \cite{BR2011}. 
When $v_\text{so}^\pm\neq0$, $Q$ also mediates the angular momentum leakage from the DQD due to the spin-orbit coupling.  Because the integral for $P$ converges fast at the scale of $t\sim1/|v^\pm|$ while the integral for $Q$ diverges as $\ln T_\text{LZ}$ for linear LZ sweeps, $Q$ is typically large, especially for $P\approx1$, and deeply influences the self-quenching process.

Using the amplitudes $(c_S(t),c_{T_+}(t))$ found from solving Eq.~(\ref{Hst}) in combination with the dynamical equations for 
nuclear spins of Eq.~(\ref{spinchanges}) makes our approach completely self-consistent. During a single LZ sweep, $\epsilon_S(t)$ and $\epsilon_{T_+}(t)$ change fast while staying close to the anticrossing, and
$\eta$ and $v^\pm$ remain practically constant. The singlet wave function of the DQD in the $S-T_+$ anticrossing point can be expressed as $\psi_S=\cos\nu~\psi_{02}+\sin\nu~\psi_{11}$, where $\psi_{02}$ and $\psi_{11}$ are the singlet wave functions with both electrons on the right dot and two electrons equally distributed between the dots, respectively. The mixing angle $\nu$ is controlled by $B$ and detuning  and was chosen as $\nu=\pi/4$.

Our simulations included about $10^7$  nuclei up to those with a hyperfine interaction strength of only 1\% of the maximum one, which allowed us to account for the electronic density heterogeneity and the spin polarization transfer from the interior to the periphery. Our algorithm allowed performing calculations for different $T_\text{LZ}$ with the identical initial distribution.

For each sweep,  
the DQD is first set in its eigenstate at $t=-T_\text{LZ}$ that is close, but not identical, to
the singlet (0,2) state with both electrons localized at the right dot. Then a change in the gate voltages drives a (partial) transition to the triplet (1,1) state with electrons shared between both dots. Finally the electronic system is reset in its initial state. We assume that the LZ transition time $T_\text{LZ}$ are much shorter than the  nuclear precession times $t_{pr}$  
and compute the change in the direction of each of the nuclear spins during every sweep numerically, as described by Eqs.~(\ref{Hst})-(\ref{eq:PandQ}). Between consecutive sweeps, repeated with a period of the waiting time $T_\text{w}$, electrons are in the singlet state and do not interact with the nuclear spins that coherently precess in the external field. We choose realistic parameters for a parabolic DQD of a height $w=3$ nm, size $\ell=50$ nm, and interdot separation $d=100$ nm; magnetic field $B$=10 mT. All the results presented below were found for the same initial configuration of nuclear spins, but we have checked that they  are representative for 
generic initial configurations.

For the shape of the {$S\rightarrow T_+$ pulses, we used the LZ model with 
$\epsilon_s(t) = \epsilon_\text{max}t/2T_\text{LZ}$ and $\epsilon_{T_+}-\eta=-(\epsilon_\text{max} t/2T_\text{LZ}) - (\eta - \eta_i)$, where $\eta_i$ is the initial polarization $\eta_i=\eta(t=-T_\text{LZ})$ and $\epsilon_\text{max}=2.5$ meV, which is larger than the typical $S$-$T_+$ coupling.
To avoid trivial quenching due to the shift in $\eta$ caused by the accumulating polarization, the electronic energies  were renormalized after every 100  
sweeps to keep $\eta - \eta_i \approx 0$. As a result, the center of the sweep was permanently kept close to the anticrossing point. Such a regime can be achieved experimentally by applying appropriate feedback loops.  SO coupling in DQDs is device specific, and in GaAs it changes from weak to moderate, so we consider it both in the limit of no SO coupling \cite{Petta2005,Brunner} and with SO coupling of a reasonable magnitude \cite{Ganichev2004,Nowack2007}. 

Fig.\ \ref{GaAsmultispecie12345}(b) plots the change in the total nuclear polarization, $\Delta I_z$, as a function of the number of sweeps  $n$, for $v_\text{so}=0$ and four transition times 
$T_\text{LZ}$; the difference in the parameter values of all three nuclear species is taken into account. The evolution of $\Delta I_z$  typically saturates within $3 \times 10^4$ sweeps. The saturation proves the self-quenching of the transverse nuclear polarizations that controls the singlet-triplet coupling $v_\text{n}^\pm$ shown in Fig.\ \ref{GaAsmultispecie12345}(c). It vanishes after a number of LZ transitions and this typically happens faster for longer LZ  
durations $T_\text{LZ}$ (a larger transition probability), but more complicated patterns of subsequent revivals of the $v_\text{n}^\pm$ can also be seen.  Volatile dynamics of $v_\text{n}^\pm$ seen in Fig.\ \ref{GaAsmultispecie12345}(c), with multiple maxima and minima, is typical of multi-specie systems because of the different spin precession rates of different species. However, finally the nuclear subsystem self-synchronizes in one of the states in which it decouples from the electron spin qubit, and this is our first {\it central result}. 
The contribution of each of the species to $v_\text{n}^\pm$ vanishes identically, at each instant; $v_\text{n}^\pm=0$ persists even after the LZ pumping is interrupted (not shown). This result resembles the ``dark states" of Refs.~\onlinecite{TaylorIL:PRL2003,Gullans:prl10}.

With $N\sim10^6$ nuclei in the DQD, the initial fluctuations producing $v_\text{n}^\pm$ is 
$N^{1/2}\sim10^3$. For $P\sim1$, one expects that at least $n\sim10^3$ pulses are needed for balancing it. The typical number of pulses to establish self-quenching of about $n\sim10^4$ of Fig.~\ref{GaAsmultispecie12345}(b) is one order of magnitude larger,  
which can be attributed to the high volatility of the process and the fact that the LZ probability $P$ in each cycle is less than one. 
\begin{figure}[htbp]
\includegraphics[width=0.9\columnwidth]{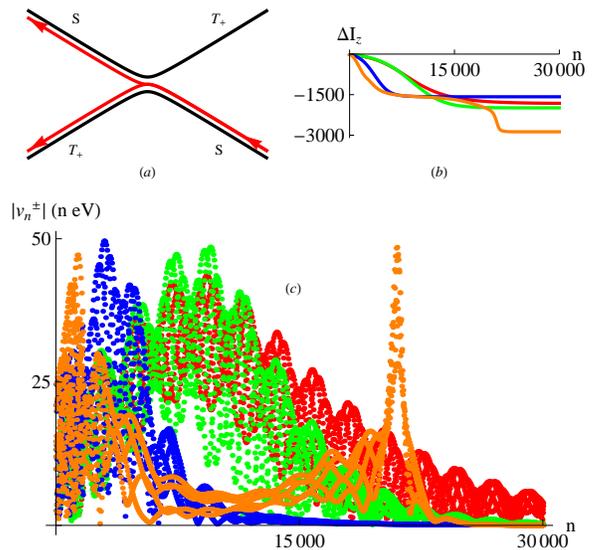}
\caption{Nuclear dynamics in absence of SO coupling as a function of 
sweep number $n$; difference in the parameters of the three nuclear species is taken into account. Resonant waiting time  
$T_\text{w}=t_\text{75As}=13.7 \mu s$; qualitative pattern do not depend on this specific choice.  (a) Landau-Zener passage of a singlet $S$ through a $S-T_+$ anticrossing. Energy levels (black) and evolution of $S$ into entangled $S$ and $T_+$ states (red). (b) Change in the nuclear spin polarization $\Delta I_z$  and (c)  hyperfine-induced singlet-triplet coupling $|v_\text{n}^\pm|$.  Color codes: $T_\text{LZ}=10$ ns (red), 20 ns (green), 40 ns (blue), and 80 ns (orange). 
}
\label{GaAsmultispecie12345}
\end{figure}

Next, we consider the effect of the SO coupling. To illustrate the main qualitative result, we ascribe to all nuclei identical parameters found by averaging over the three GaAs species \cite{Supplemental} 
and consider strictly resonant pumping, $T_\text{w}=t_\text{GaAs}=10.7 \mu s$, 
 $t_\text{GaAs}$ being the nuclear precession time \cite{multi}.
We use realistic values of the SO coupling, $v_\text{so}=0$, $31,$ and $62$ neV \cite{SO}
(and choose it to be a real number). The LZ transition probability shown in  Fig.~\ref{GaAssinglespecie12345Tp160ns}(a) vanishes at large $n$, hence, the pumping is self-quenched also in this case. At this point, it is crucial to note that all our numerical results for 
$v_\text{n}^\pm$ and $\Delta I_z$ are plotted at multiples of the waiting time $T_\text{w}$.  Hence,  self-quenching in a resonant ($T_\text{w}=t_\text{GaAs}$) SO coupled system is achieved through $v_\text{n}^\pm \rightarrow -v_\text{so}$  ($v^\pm = v_\text{n}^\pm + v_\text{so}^\pm$) at every multiple 
of the Larmor period, and nuclear polarization screens the SO  coupling, 
Fig.~\ref{GaAssinglespecie12345Tp160ns}. This screening of SO coupling is out next {\it central result}. However, in contrast to the case of no SO coupling, between the sweeps (not shown), the matrix elements $v_\text{n}^\pm(t)$ change harmonically with the amplitude $v_\text{so}$ and a period $t_{GaAs}$, $v_\text{n} \rightarrow v_\text{so} \cos{(2 \pi t/t_\text{GaAs})}$, where the time $t=n T_\text{w}=nt_\text{GaAs}$ at the LZ sweep number $n$. Not surprisingly, while $\vert v^\pm_\text{n}\vert$ reaches its $T_\text{LZ}$-independent limit, the change in the polarization, $\Delta I_z$, depends on $T_\text{LZ}$.  Fig.~\ref{GaAssinglespecie12345Tp160ns}(a) shows that $P$ is large, $P\sim1$, and fluctuates fast with $n$. The mechanism of fast dynamics is unveiled by the high magnitude of the shakeup parameter $Q\sim20\gg 1$. While $P$ describes pure injection of the angular momentum, $Q$ describes its
redistribution due to shakeup processes and the SO coupling \cite{BR2011}. Remarkably, it is seen from Eq.~(\ref{DeltaIz}) 
that the effect of the $Q$ term on $\Delta I_z$ vanishes in two important limits, when $v_\text{so}=0$ and $v_\text{n}^\pm=-v_\text{so}$.   Fig.~3(c) demonstrates that self-quenching sets in 
sharply for large $T_\text{LZ}$, and the fluctuational phase lasts longer for larger $v_\text{so}$ values; $v_\text{n}^\pm$ always saturate 
at $-v_\text{so}$. For SO coupled systems, the
change $\Delta I_z$ in the 
magnetization is non-monotonic in $n$ and shows no regular dependence on $v_\text{so}$, Fig.~\ref{GaAssinglespecie12345Tp160ns}(d).
\begin{figure}[htbp]
\includegraphics[width=0.9\columnwidth]{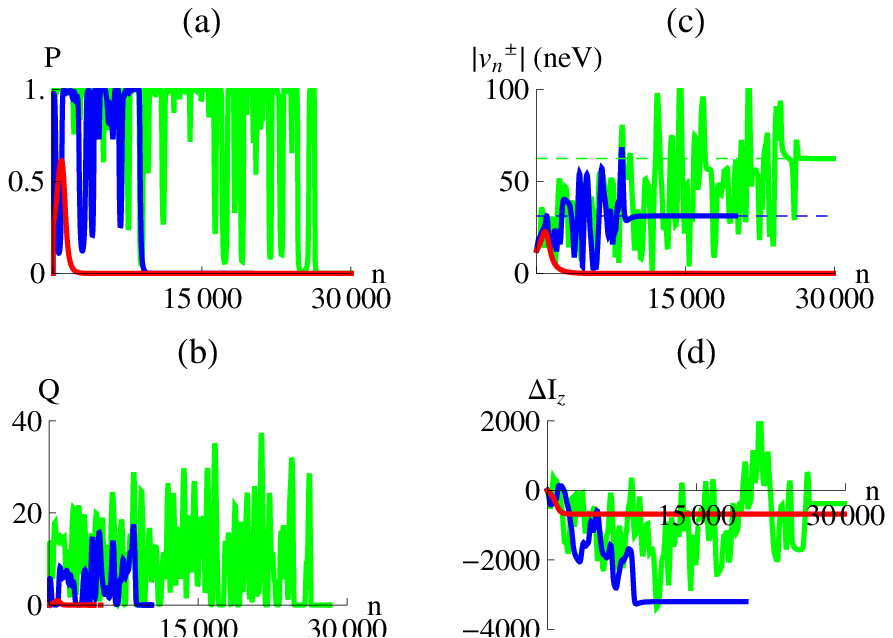}
\caption{} Nuclear dynamics for variable spin-orbit coupling $v_\text{so}$ driven by long pulses $T_\text{LZ}=160$ ns; single-specie model {with $T_\text{w}=t_{GaAs}$}. (a) Landau-Zener probability $P$, (b) shakeup parameter $Q$, (c) hyperfine-induced singlet-triplet coupling $\vert v_\text{n}^\pm\vert$, and (d) change $\Delta I_z$ in the total nuclear polarization; in (c), dashed lines mark spin-orbit coupling. Color codes: $v_\text{so}=0$ (red), 31 neV (blue), and 62 neV (green). 
\label{GaAssinglespecie12345Tp160ns}
\end{figure}

To test how robust
our results are, we modeled the influence of noise by adding a random magnetic field along the $z$-direction for each nuclear spin so that the nuclear spins acquire an additional phase of $2 \pi r_{j \lambda}T_\text{w}/\tau$ during each waiting time  
between LZ sweeps, where $r_{j\lambda}$ are random numbers in the interval from -1 to 1; $\tau$ is the noise correlation time (see Ref.~\cite{Supplemental}) for more details).

Fig.~\ref{GaAssinglespecie12345Noise} demonstrates the effect of the noise that randomizes the phases of the precessing nuclear spins.  
It displays  the magnitude of the hyperfine matrix element $v_\text{n}^{\pm}$ at every multiple of the waiting time $T_\text{w}$} for no noise and for two noise levels. 
In all cases, $ v_\text{n}^\pm$ approaches the value $\approx-v_\text{so}$ at $n\alt50000$.  While in the absence of the noise the saturation of $v_{\text n}^\pm$ is exact at all multiples of the waiting time
$T_\text{w}$, $v_\text{n}^\pm=-v_\text{so}$ (blue curve), noisy systems experience slight fluctuations near this value, see especially the red line which only saturates 
at $n\approx 5 \times 10^4$. 
We conclude that for moderate noise levels, the dynamical screening is  
robust, and this is our final {\it central result}. 
\begin{figure}[htbp]
\includegraphics[width=0.9\columnwidth]{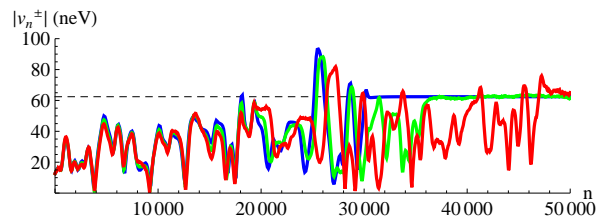}
\label{GaAssinglespecie12345noise80_62}
\caption{Effect of noise on nuclear dynamics. The LZ sweep duration of $T_\text{LZ}=80$ns; single-specie model with $T_\text{w}=t_\text{GaAs}$. Hyperfine-induced singlet-triplet coupling 
 $|v_\text{n}^\pm|$ for spin-orbit coupling $v_\text{so}=62~ n$eV  and increasing levels of transverse noise. Black dashed lines show SO coupling. Color codes: $\tau/t_\text{GaAs}=\infty$ (blue), 5000 (green), and 2500 (red); $t_\text{GaAs}$ is the nuclear spin precession time and $\tau$ is the noise correlation time.
} 
\label{GaAssinglespecie12345Noise}
\end{figure}

The above results prove that dynamical self-quenching is a rather generic property of the GaAs type DQDs pumped by multiple passages through the $S - T_+$ anticrossing \cite{SOcouplNR}. In a quenched state, the electron spin qubit becomes screened 
from the randomness of the nuclear spin bath, and therefore its decoherence  by nuclei \cite{Yao2006,Cywinsky2009,Fisher2009} is expected to be
suppressed; trapping the qubit into the quenched state can be checked by spin splitter technique \cite{Petta2010,Ribeiro2010}. A quenched qubit can be operated by short pulses applied to the gates due to the different dependence of $v_\text{n}^\pm$ and $v_\text{so}$ on the shape of the electron wave functions. This subject requires further
consideration.

In conclusion, self-quenched states produced by stationary pumping of a spin blockaded double quantum dot by multiple passages through the $S-T_+$ anticrossing decouples
the electronic qubit from the nuclear spin bath. In such states the dynamical nuclear polarization screens both the initial random Overhauser field and a moderate spin orbit coupling typical of GaAs quantum dots. Self-quenching unveils the origin of the low spin pumping efficiency encountered in experimental studies.

A.\ B.\ would like to thank B.\ I.\ Halperin for his hospitality at Harvard University where this work was initiated. E.\ I.\ R.\ acknowledges funding from the  Intelligence Advanced Research Project Activity (IARPA) through the Army Research Office.


\pagebreak\newpage

\begin{widetext}

\section{EPAPS Supplementary Material for "Dynamical Self-Quenching of Spin Pumping into Double Quantum Dots"}

In this supplementary, we provide details of the parameters used in the numerical calculations, describe the effective single-specie model, show how nuclear spins precess in an external magnetic field between Landau-Zener sweeps, and discuss the noise model.

\subsection{Parameters}

GaAs has  8 nuclear spins per cubic unit cell so that the effective volume per site is $V_{s}=a^{3}/8$, where the lattice constant is $a=5.65$\AA  .   When all nuclear spins are fully polarized, the Overhauser field seen by the electrons is $5.3$ T. The electron $g$-factor is $g_\text{GaAs}=-0.44$. The other parameters reflecting the abundance, $g$-factors, and hyperfine coupling constants are listed in Table \ref{speciesvalues}. 
\begin{table}[h]
\begin{center}
\begin{tabular}{|r|r|r|r|}
\hline
 & 69Ga & 71Ga & 75As  \\
\hline
$p$ & 30\% & 20\% & 50\% \\
\hline
$g$ & 1.3 & 1.7 & 0.96 \\
\hline
$A$ ($\mu$eV)& 77 & 99 & 94 \\
\hline
$I$ & 3/2 & 3/2 & 3/2 \\
\hline
\end{tabular}
\caption{\label{speciesvalues} Nuclear abundances $p$, nuclear $g$-factors, hyperfine coupling constants $A$, and nuclear spin in GaAs.}
\end{center}
\vspace{-0.6cm}
\end{table}

\subsection{Effective Single-Specie Model}

To illustrate the main qualitative effects of SO coupling, we also make use of an effective single-specie model. In the effective single-specie model, the mean nuclear $g$-factor is $g_\text{GaAs}=\sum_\lambda p_\lambda g_\lambda=1.21$, the mean hyperfine coupling constant is $A_\text{GaAs}=\sum_\lambda p_\lambda A_\lambda= 90 \mu$eV, and the mean nuclear spin is $I_\text{GaAs}=\sum_\lambda p_\lambda I_\lambda = 3/2$.

\subsection{Nuclear Precession in the External Magnetic Field}

Landau-Zener sweeps of a duration $T_\text{LZ}$ are repeated with with wait periods of 
$T_\text{w}$ long compared with $T_\text{LZ}$, hence, $T_\text{LZ} \ll T_\text{w}$.  Between consecutive Landau-Zener sweeps, electrons are in the singlet state and do not interact with the nuclear spins. However, during the waiting
 interval $T_\text{w}$, nuclear spins precess in an external magnetic field $B$ applied along the $z$-direction. Then, the coherent evolution of the nuclear spins are 
\begin{subequations}
\begin{align}
I_{j\lambda}^x(t+T_\text{w}) & = I_{j\lambda}^x(t) + I_{j\lambda}^x(t) \cos{\phi_{j\lambda}}  - I_{j\lambda}^y(t) \sin{\phi_{j\lambda}} , \\
I_{j\lambda}^y(t+T_\text{w}) & = I_{j\lambda}^y(t) + I_{j\lambda}^y(t)  \cos{\phi_{j\lambda}} +I_{j\lambda}^x(t) \sin{\phi_{j\lambda}}  , \\
I_{j\lambda}^z(t+T_\text{w}) & = I_{j\lambda}(t) ,
\end{align}
\label{spinprecession}
\end{subequations}
where $t$ is the initial time of the interval, the transverse phase change is $\phi_{j \lambda}=-2 \pi T_\text{w}/t_\lambda$ in terms of the spin precession time $t_\lambda=2 \pi \hbar/g_\lambda \mu_I B$, $g_\lambda=\mu_\lambda/I_\lambda$ is the $g$-factor for a nuclear specie $\lambda$, $\mu_\lambda$ its magnetic moment and $\mu_I=3.15 \times 10^{-8}$ eV/T is the nuclear magneton.

\subsection{Noise}

We neglected in our calculations the effects of the magnetic dipole-dipole interaction between the nuclear spins that is slow at our time scale and also the possible electrical noise on the gates . The detailed study of these effects is beyond the scope of our paper. 

We use a simple model to consider the influence of such noise by phenomenologically adding a random magnetic field along the $z$-direction for each \emph{individual} nuclear spin so that the accumulated phase in Eq.~\ref{spinprecession} changes,
$\phi_{j\lambda} \rightarrow\phi_{j\lambda}^\text{eff}=\phi_{j\lambda}+\phi_{j\lambda}^\text{noise}=-2 \pi T_\text{w} (1/t_\lambda - r_{j\lambda}/\tau)$,  
where $r_{j\lambda}$ are random numbers in the interval from $-1$ to $1$ \emph{for each nuclear spin}. This procedure simulates a randomization of the transverse components of the nuclear spins after a time of the order $\tau$, and $\tau$ 
is termed the noise correlation time. 

Quantitively, the noise averages of the phase-dependent factors in Eq.~\ref{spinprecession} become
\begin{align}
\frac{1}{2} \int_{-1}^1 dr \cos{\phi_{j\lambda}^\text{eff}} &= \cos{\phi_{j\lambda}} \frac{\sin{(2 \pi T_\text{w}/\tau})}{2 \pi T_\text{w}/\tau} \, , \\
\frac{1}{2} \int_{-1}^1 dr \sin{\phi_{j\lambda}^\text{eff}} & = \sin{\phi_{j\lambda}} \frac{\sin{(2 \pi T_\text{w}/\tau})}{2 \pi T_\text{w}/\tau} \, , 
\end{align}
so that the phase-coherent information contained in the transverse nuclear spins components are lost after a time of the order of the noise correlation time $\tau$. By introducing noise, any possible build-up of the matrix elements $v_\text{n}^\pm$ is therefore lost after a time of the order of the noise correlation time $\tau$.

\end{widetext}

\end{document}